# UT1 prediction based on long-time series analysis

**Tissen V.M., Tolstikov A.S.,** Siberian Scientific Research Institute of Metrology, Novosibirsk, Russia

**Malkin Z.M.,** Pulkovo Observatory, St. Petersburg, Russia

*Abstract. A new method is developed for prediction of UT1. The method is based on construction of a general harmonic model of the Earth rotation using all the data available for the last 80-100 years, and modified autoregression technique. A rigorous comparison of UT1 predictions computed at SNIIM with the prediction computed by IERS (USNO) in 2008-2009 has shown that proposed method provides substantially better accuracy.*

## Introduction

EOP prediction is one of the most important parts of EOP service activity. As practical requirements to the EOP prediction accuracy are toughen, methods of EOP prediction should be regularly improved. In this paper we present a new method of prediction of UT1 developed at the Siberian Scientific Research Institute of Metrology (SNIIM). This method is being developed at SNIIM during last several years, and since 1997 is thoroughly tested at the Pulkovo Observatory to provide an independent estimate of the its real accuracy. Below the brief description of this method and test results are presented.

## Method

The main distinctive of our method is making use of long-time series up to 100 years to estimate the trend component of the UT1 series. This trend component is expanded in a harmonic time series consisting of 20 and more terms with periods from 66 to 1/3 years. Main trend harmonic components are shown in Table 1.

Table 1. Main trend components.

| Period, years | Amplitude, ms | Period, years | Amplitude, ms | Period, years | Amplitude, ms | Period, years | Amplitude, ms |
|---|---|---|---|---|---|---|---|
| 66 | 3500 | 8 | 20 | 2.0 | 4 | 0.83 | 1.2 |
| 33 | 500 | 6 | 40 | 1.67 | 4 | 0.71 | 1.1 |
| 22 | 750 | 4.8 | 25 | 1.10 | 2 | 0.58 | 1 |
| 14 | 160 | 3.6 | 15 | 1.0 | 21 | 0.50 | 8 |
| 10 | 50 | 2.4 | 15 | 0.9 | 1 | 0.33 | 1 |

After removing trend, the residuals are predicted making use of a modified autoregression method. Before this procedure tidal and some other known regular variations in UT1 are removed, and then added back to form the final prediction.

**Testing**

The method proposed was tested by means of comparison with the IERS Rapid Service/Prediction Centre (USNO) predictions. To provide such a comparison, IERS operational series computed at USNO was extrapolated at SNIIM. One or two 90-day predictions a week were computed, 156 predictions in total from 10 Jan 2008 till 07 Oct 2009. We use 90-day prediction for this analysis because, unfortunately, USNO provides only such a length of prediction for the daily solutions. Our predictions were stored together with the IERS ones made on the same day. Thus, we collected a set of 156 pairs of predictions made on the same day at SNIIM and USNO using the same observed EOP series. Afterwards these predictions were compared to the final USNO EOP series, and prediction errors were computed. To provide a more detailed comparison, three statistics were used: RMS error, MAE and maximum error.

**Results**

Results of processing of the series of predictions described above are presented in Figs. 1 and 2 for 90-day and 7-day respectively. One can see that the method presented show better accuracy, especially for shortest (up to 7-10 days) and longest (>30 days) predictions. It should be mentioned that the method details are regularly improved, and one can see that the results obtained for 2009 are substantially better than ones for the whole interval 2008-2009, whereas USNO results for 2009 are a bit worse that ones for the whole interval.

**Conclusion**

In this paper a new method of UT1 prediction developed at SNIIM is presented. It was rigorously compared with the IERS results using 156 predictions made in 2008-2009 simultaneously at SNIIM and USNO. Method proposed here has shown the better accuracy, and prospects for further improvement are optimistic. Especially important for GNSS applications is an improvement in the short-time prediction with the length of several days. As to more long-time prediction, 30 days and more, it was shown that using long-time reference series for trend evaluation can provide better prediction accuracy.

More details of this study can be found in *(Tissen et al., 2009)*.

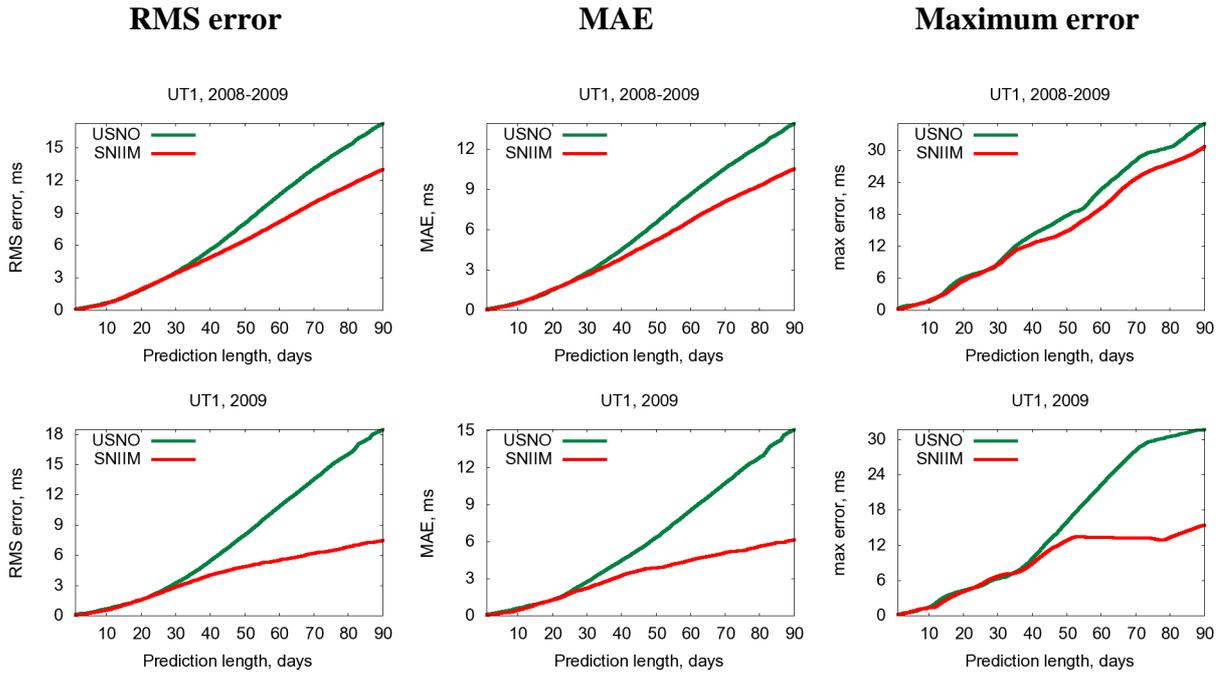

Fig. 1. UT1 90-day predictions made at SNIIM and USNO in 2008-2009 (top row) and in 2009 only (bottom row).

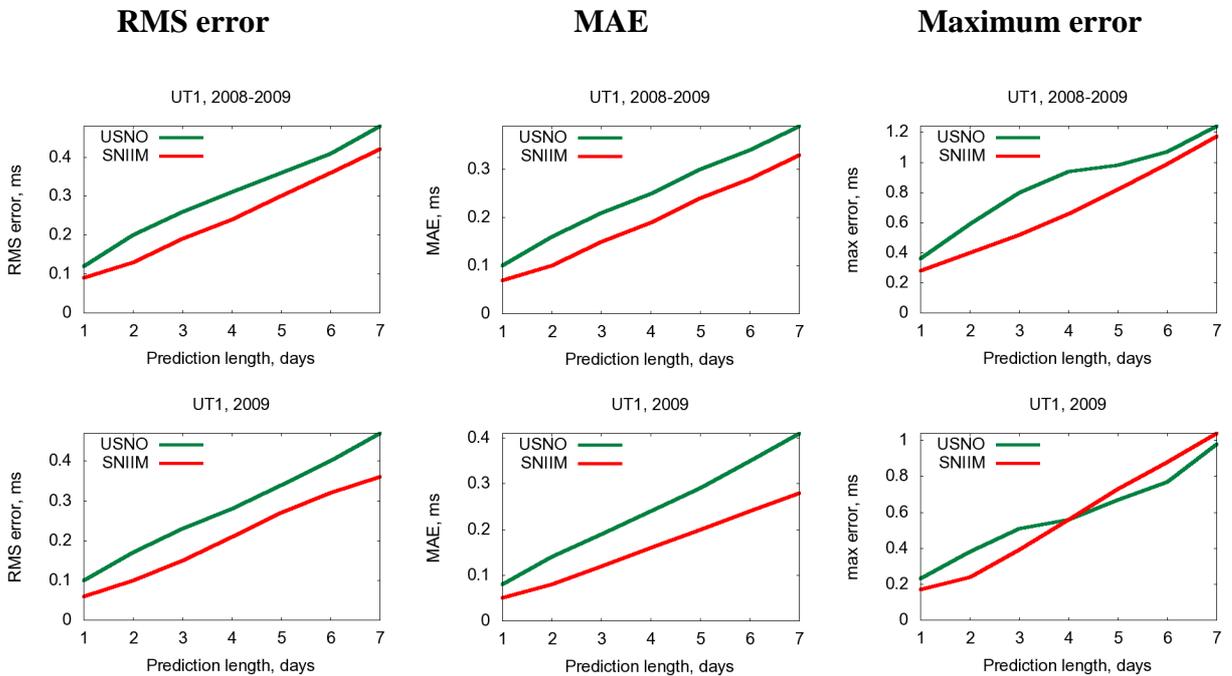

Fig. 2. UT1 7-day predictions made at SNIIM and USNO in 2008-2009 (top row) and in 2009 only (bottom row).